\documentclass[reprint, superscriptaddress, amsmath, amssymb, aps, prl floatfix]{revtex4-2}
\usepackage{graphicx} 
\usepackage{bm} 
\usepackage{mathtools} 

\usepackage[caption=false]{subfig}
\usepackage[usenames,dvipsnames]{color}
\usepackage{xcolor}
\usepackage{orcidlink}
\usepackage{booktabs}

\newcommand{\ICTS}{International Centre for Theoretical Sciences,
    Tata Institute of Fundamental Research, Bangalore 560089, India}
\newcommand{\IITM}{Department of Physics, Indian Institute of Technology Madras, Chennai 600036, India}
\newcommand{\CSGC}{Centre for Strings, Gravitation and Cosmology, Department of Physics, Indian Institute of Technology Madras, Chennai 600036, India}
\newcommand{\AEI}{Max Planck Institute for Gravitational Physics (Albert Einstein Institute), Am M{\"u}hlenberg 1, 14476 Potsdam, Germany}
\newcommand{\UMass}{Department of Mathematics, Center for Scientific Computing and Data Science Research, University of Massachusetts, Dartmouth, MA 02747, USA
}

\newcommand{\param}{\bm{\theta}}
\newcommand{\eref}{e_{\rm{ref}}}
\newcommand{\lref}{l_{\rm{ref}}}

\newcommand{\esigma}{\texttt{ESIGMA}}

\newcommand{\inspiralesigma}{\texttt{InspiralESIGMA}}
\newcommand{\inspiralesigmasur}{\texttt{InspiralESIGMASur}}
\newcommand{\imresigma}{\texttt{IMRESIGMA}}

\newcommand{\esigmapy}{\texttt{ESIGMAPy}}

\begin{document}

\title{Chase Orbits, not Time: A Scalable Paradigm for Long-Duration Eccentric Gravitational-Wave Surrogates}

\author{Akash Maurya \orcidlink{0009-0006-9399-9168}}\email{akash.maurya@icts.res.in}\affiliation{\ICTS}
\author{Prayush Kumar \orcidlink{0000-0001-5523-4603}}\email{prayush@icts.res.in}\affiliation{\ICTS}
\author{Scott E. Field \orcidlink{0000-0002-6037-3277}}\affiliation{\UMass}
\author{Chandra Kant Mishra \orcidlink{0000-0002-8115-8728}}\affiliation{\CSGC}\affiliation{\IITM}
\author{Peter James Nee \orcidlink{0000-0002-2362-5420}}\affiliation{\AEI}
\author{Kaushik Paul \orcidlink{0000-0002-8406-6503}}\affiliation{\ICTS}\affiliation{\CSGC}\affiliation{\IITM}
\author{Harald~P.~Pfeiffer~\orcidlink{0000-0001-9288-519X}}\affiliation{\AEI}
\author{Adhrit Ravichandran~\orcidlink{0000-0002-9589-3168}}\affiliation{\UMass}
\author{Vijay~Varma~\orcidlink{0000-0002-9994-1761}}\affiliation{\UMass}

\begin{abstract}
    Orbital eccentricity is a key tracer of the astrophysical origins of compact binaries. Yet it remains absent from routine LIGO--Virgo--KAGRA analyses, in part because of the prohibitive computational cost of generating eccentric template waveforms. The complicated morphology of these waveforms due to the eccentric orbital timescale variations makes it difficult to construct their accurate and efficient surrogate models, especially for waveforms long enough to comprehensively cover the sensitivity bands of current ground-based gravitational-wave detectors. We present a novel and scalable surrogate building technique which makes surrogate modeling of long-duration eccentric binary black hole waveforms both feasible and highly efficient. The technique aims to simplify the harmonic content of intermediate eccentric waveform data-pieces by modeling them in terms of an angular orbital element called the mean anomaly, instead of time. We show that this parameterization yields much more compressed surrogates than the standard time-based parameterizations. 
    We also significantly simplify variations in waveform data-pieces across the parameter space by expressing them in terms of the instantaneous orbital eccentricity and mean anomaly to ease their parametric fitting.
    Building on these developments, we construct \texttt{InspiralESIGMASur}: a $2.77 \times 10^6M$ (850-1250 orbits) long non-spinning surrogate for the inspiral-only eccentric waveform model \texttt{InspiralESIGMA} [K. Paul \textit{et al.}, \href{https://doi.org/10.1103/PhysRevD.111.084074}{Phys. Rev. D 111, 084074 (2025)}]. The methods presented in this work make it feasible to build long-duration eccentric surrogates for current as well as future third-generation gravitational-wave detectors.
\end{abstract}

\date{July 21, 2026}

\maketitle

Since the first detections of gravitational waves (GWs) from compact binary coalescences (CBCs) by the LIGO--Virgo--KAGRA (LVK) collaboration~\citep{LIGOScientific:2018mvr, LIGOScientific:2020ibl, KAGRA:2021vkt, LIGOScientific:2025slb}, the formation channels and host environments of these binaries have remained uncertain~\citep{LIGOScientific:2018jsj, LIGOScientific:2020kqk, KAGRA:2021duu, LIGOScientific:2025pvj}.
Binaries formed in isolation are expected to circularize efficiently through GW emission~\citep{Postnov:2014tza, PhysRev.136.B1224}, whereas binaries assembled in dense stellar environments may retain measurable eccentricity due to dynamical interactions with other bodies~\citep{Samsing:2013kua, Antonini:2015zsa, Samsing:2017rat, Rodriguez:2017pec, Zevin:2018kzq, Samsing:2020tda, Fabj:2024jns, Samsing:2024syt}.
This makes orbital eccentricity a key tracer of binary formation channels. There already exist claims of eccentric detections, though they remain under active investigation~\citep{Wu:2020zwr, Romero-Shaw:2021ual, Romero-Shaw:2022xko, Romero-Shaw:2020thy, Gayathri:2020coq, Gupte:2024jfe, Planas:2025jny, Morras:2025xfu, Planas:2025plq, Kacanja:2025kpr, Jan:2025fps}. 

GW data analysis is computationally demanding: matched-filter CBC searches and Bayesian parameter estimation typically require generating $\mathcal{O}(10^{6+})$ template GW waveforms. Although many inspiral-only~\citep{Yunes:2009yz, Cornish:2010cd, Huerta:2014eca, Tanay:2016zog, Tiwari:2019jtz, Tiwari:2020hsu, Moore:2018kvz, Moore:2019xkm, Sridhar:2024zms, Klein:2018ybm, Klein:2021jtd, Morras:2025nlp} and inspiral-merger-ringdown (IMR) eccentric waveform models~\citep{Paul:2024ujx, Hinder:2017sxy, Liu:2021pkr, Liu:2023dgl, Gamba:2024cvy, Nagar:2024dzj, Gamboa:2024hli, Planas:2025feq, Chattaraj:2022tay, Islam:2024zqo, Manna:2024ycx} now exist, most models rely on expensive numerical solving of coupled ordinary differential equations (ODEs) for the orbital dynamics, limiting their use in large-scale inference. This has hindered eccentric parameter estimation via standard approaches~\citep{Romero-Shaw:2021ual, Romero-Shaw:2022xko, Romero-Shaw:2020thy} and has motivated alternative inference methodologies~\citep{Gupte:2024jfe}.

Surrogate and reduced-order modeling (ROM) techniques have proven invaluable for accelerating expensive \textit{non-eccentric} waveform models by constructing their fast, high-fidelity approximations via dimensionality/complexity reduction~\citep{Field:2013cfa, Purrer:2014fza, Chua:2018woh, Lackey:2018zvw, Cotesta:2020qhw, Khan:2020fso, Gadre:2022sed, Thomas:2022rmc, Blackman:2015pia, Blackman:2017pcm, Varma:2018mmi, Varma:2019csw, Williams:2019vub, Yoo:2023spi}, and are now standard in GW data analysis~\citep{LIGOScientific:2018mvr, LIGOScientific:2020ibl, KAGRA:2021vkt, LIGOScientific:2025slb}. However, surrogate modeling of \textit{eccentric} systems has remained challenging~\citep{Islam:2021mha, Yun:2021jnh, Shi:2024age, Islam:2025llx, Islam:2025rjl} due to their intricate waveform morphology. This issue becomes particularly severe for long-duration signals, and to our knowledge, no existing methodology is capable of fully addressing it. The problem will become increasingly acute for the next-generation (3G) detectors~\citep{2017arXiv170200786A, Punturo:2010zz, Hild:2011np, Reitze:2019iox, LIGOScientific:2016wof}. They will observe significantly ($10-100 \times$) longer inspirals, probing binaries early in their evolution with potentially sizable orbital eccentricity~\citep{Saini:2023wdk}, and at substantially higher detection rates, demanding even longer waveforms and faster analyses~\citep{Couvares:2021ajn, LISAConsortiumWaveformWorkingGroup:2023arg}.

In this \textit{Letter}, we present a novel paradigm to facilitate highly scalable, accurate, and efficient surrogate construction for long-duration eccentric binary black hole (BBH) waveforms. The central idea is to simplify the harmonic content of intermediate waveform data-pieces by modeling them against the orbital \textit{mean anomaly}~\citep{goldstein2002classical, Memmesheimer:2004cv}, a kinematic parameter which {\it advances uniformly with the binary's orbits}, instead of time. We show that surrogate models built from these orbit-parameterized waveform data-pieces achieve significantly higher data compression than their time-parameterized counterparts, allowing scalable construction of long-duration surrogates.

We demonstrate this by constructing \inspiralesigmasur{}: a $2.77 \times 10^6M$ (850-1250 orbits) long non-spinning surrogate for the inspiral-only eccentric waveform model \inspiralesigma{}~\citep{Paul:2024ujx, esigmapy}. It can be used to analyze GW data from a frequency of $15$Hz from black hole binaries with total masses $M \geq 8.3M_\odot$, mass-ratio $q \coloneq m_1/m_2 \in [1,6]$, and orbital eccentricity $e_{\rm{ref}} \in [0, 0.43]$.
\inspiralesigmasur{} is publicly available through the Python package \esigmapy{}~\citep{esigmapy, esigmasur-zenodo}. 
This formulation scales well to long-duration eccentric surrogates, which makes it suitable for current as well as the future third-generation gravitational-wave detectors.


\textit{Setting up the problem---} 
Surrogate modeling exploits the simplicity and similarity in waveform features to capture the entire training waveform space with only a few representative \textit{basis functions}, thus producing a significantly compressed and efficient waveform model~\citep{Field:2013cfa}. It is therefore standard to decompose waveforms into \textit{waveform data-pieces} with simpler features, construct \textit{their} surrogates, and recombine them to produce the final surrogate waveforms. \textit{The aim of this work is to develop a paradigm for eccentric waveform data-pieces that enables their scalable and efficient surrogate modeling.}

We build surrogates using the greedy basis and empirical interpolation method via the Python package \texttt{RomPy}~\citep{rompy, Field:2013cfa}.
Unless stated otherwise, we shall use geometric units, where $G=c=1$.


\textit{Waveforms and alignment---}
We use the piecewise-IMR waveform model \imresigma{}~\citep{Paul:2024ujx, esigmapy} for eccentric BBHs, which comprises of a long-duration computationally expensive inspiral piece, \inspiralesigma{}, connected to a short merger-ringdown piece a few orbits before merger. We construct non-spinning surrogates for the inspiral $(2,2)$-mode piece \inspiralesigma{} here. Each \inspiralesigma{} waveform ends at a common orbit-averaged orbital frequency, and we set this time as $t=0$. We also fix the starting $(2,2)$-mode phase of each waveform to zero, i.e. $\phi_{22}(t=-T) = 0$ (cf. Eq. \eqref{eqn: amp-phase-decomp}), where $T$ is the time-duration of the waveform.

\begin{figure}
    \centering
    \includegraphics[width=\columnwidth, trim=0.25cm 0.275cm 0.25cm 0.25cm, clip]{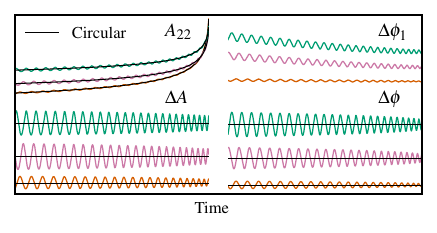}
    \caption{Amplitude $A_{22}$, eccentric residual amplitude $\Delta A$, residual phase $\Delta \phi_1$, and detrended residual phase $\Delta \phi$ (cf. Eq. \eqref{eqn: amp-phase-decomp}--\eqref{eqn: residual-circular-phase}) for a few representative $20,000M$ long eccentric \inspiralesigma{} waveforms. 
    The corresponding quasi-circular amplitudes are also shown as solid black lines.
    }
    \label{fig: eccentric-residuals}
\end{figure}

\textit{Waveform data-pieces---}
We first decompose the $(2,2)$-mode into amplitude ($A_{22}$) and phase ($\phi_{22}$), i.e.,
\begin{equation}
    h_{22}(t; \param) = A_{22}(t; \param)\, \exp{(\mathrm{i} \phi_{22}(t; \param))},
    \label{eqn: amp-phase-decomp}
\end{equation}
where $\param$ denotes the intrinsic binary parameters, namely mass-ratio ($q \coloneq m_1/m_2$), orbital eccentricity ($\eref$), and mean anomaly ($\lref$) measured \textit{at some fixed reference time} $t_{\rm{ref}}$ near the waveform's beginning. 
Eccentricity introduces orbital timescale oscillations in these amplitudes and phases, as illustrated in Fig.~\ref{fig: eccentric-residuals}. We isolate these oscillations by removing the corresponding quasi-circular quantities to get the \textit{eccentric residuals} of the amplitude and phase~\citep{Islam:2021mha}, 
\begin{align}
    \label{eqn: residual-amplitude}
    \Delta A(t; \param) &= A_{22}(t; \param) - A_{22}(t; \eref=\lref=0, \bm{\theta'})\\
    \label{eqn: residual-phase}
    \Delta \phi_1(t; \param) &= \phi_{22}(t; \param) - \phi_{22}(t; \eref=\lref=0, \bm{\theta'})
\end{align}
where $\bm{\theta'}$ denotes all the intrinsic parameters except eccentricity ($\eref$) and mean anomaly ($\lref$). This separation of eccentric oscillations becomes especially necessary for phases, for which the quasi-circular trend is quite dominant and otherwise masks them. We find that a residual monotonic trend still survives in $\Delta \phi_1$ (see Fig.~\ref{fig: eccentric-residuals}). So, we build interpolants through all the local maxima and minima of $\Delta \phi_1(t)$ and find the residual monotonic trend $\phi_{\rm{res}}(t)$ by taking their mean, and remove it from $\Delta \phi_1(t)$ to get the detrended residual phase
\begin{equation}
    \Delta \phi(t; \param) \coloneq \Delta \phi_1(t; \param) - \phi_{\rm{res}}(t; \param).
        \label{eqn: residual-circular-phase}
\end{equation}
Finally, we make surrogates for the quasi-circular amplitude $A_{22}(t; \eref=\lref=0, \bm{\theta'})$ and phase $\phi_{22}(t; \eref=\lref=0, \bm{\theta'})$, and for the eccentric residuals $\{\Delta A(t; \param), \Delta \phi(t;\param), \phi_{\rm{res}} (t;\param)\}$. The surrogate model for the eccentric $(2,2)$-mode $h_{22}(t; \param)$ can then be assembled via Eq. \eqref{eqn: residual-amplitude}, \eqref{eqn: residual-phase}, \eqref{eqn: residual-circular-phase} and Eq. \eqref{eqn: amp-phase-decomp}.

\begin{figure}
    \centering
    \includegraphics[width=\columnwidth, trim=0.35cm 0.35cm 0.32cm 0.35cm, clip]{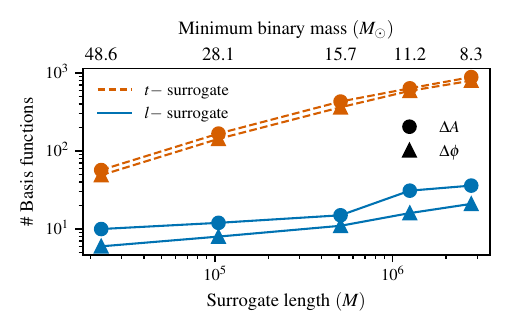}
    \caption{Number of basis functions required to achieve a basis error threshold~\citep{Field:2013cfa} of $10^{-5}$ for representing the training spaces of the eccentric residual amplitude $\Delta A$ and phase $\Delta \phi$ for different surrogate lengths for time-parameterized (dashed orange) and mean anomaly-parameterized (solid blue) surrogate methodologies. The mean anomaly parameterization achieves the same accuracy with an order of magnitude fewer basis functions. The minimum binary mass for which each surrogate can be evaluated from $15$Hz across its parameter space is indicated at the top, with additional metrics listed in Table~\ref{tab: surrogate-metrics}.}
    \label{fig: surrogate-scaling}
\end{figure}

\textit{Compressibility of data pieces---} 
Next, we express the training spaces of these data pieces in terms of a few representative basis functions using the greedy basis algorithm \cite{Field:2013cfa}. Figure~\ref{fig: surrogate-scaling} shows the scaling of the number of basis functions required for representing the training space of the eccentric residuals $\Delta A$ and $\Delta \phi$ for different durations of training space waveforms for a fixed (greedy) basis error threshold of $10^{-5}$~\citep{Field:2013cfa}. 
We observe that standard time-parameterized eccentric residuals require $\mathcal{O}(10^{2-3})$ basis functions. For comparison, a $6.02 \times 10^6M$ long surrogate for quasi-circular waveforms requires only $8$ basis functions in total for $A_{22}$ and $\phi_{22}$ (see Table~\ref{tab: surrogate-metrics}).  This highlights the central issue in eccentric surrogate modeling---the orbital timescale oscillations due to eccentricity in $\Delta A$ and $\Delta \phi$ prevent their efficient compression into a small set of basis functions. It is this issue that we alleviate by parameterizing $\Delta A$ and $\Delta \phi$ in terms of mean anomaly.
\begin{table*}[]
\caption{Summary of the surrogate models constructed in this work. For each model, we list its length, maximum initial eccentricity ($e_{\rm max}$), number of training waveforms ($N_{\rm train}$), and starting frequency ($f_0$) range for a $10M_\odot$ binary. The number of basis functions required to represent the eccentric residuals ($\Delta A, \Delta \phi$) and the median/maximum mismatch against the base model \inspiralesigma{} are shown for both time ($t$) and mean anomaly ($l$) parameterizations, demonstrating the improved compression and accuracy of the latter. All the eccentric surrogates cover mass ratios $q \in [1,6]$, while the time-parameterized quasi-circular surrogate (last row) is built for $q \in [1,8]$. Longer time-parameterized surrogates were not pursued fully owing to large basis sizes, and hence their mismatches are not reported.} 
\label{tab: surrogate-metrics}
\setlength{\tabcolsep}{5pt}
\begin{tabular*}{\textwidth}{@{\extracolsep{\fill}} c c c c c c c c}
\toprule
\multicolumn{1}{c}{Length} & \multicolumn{1}{c}{$e_{\rm max}$} & \multicolumn{1}{c}{$N_{\rm train}$} & \multicolumn{1}{c}{$f_0$ @ $10 M_\odot$} & \multicolumn{2}{c}{\# Basis functions $(\Delta A, \Delta \phi)$} & \multicolumn{2}{c}{Mismatch at $10 M_\odot$ (median, max)} \\
\cmidrule(lr){5-6} \cmidrule(lr){7-8}
($10^3 M$) & & & (Hz) & $t$--param. & $l$--param. & $t$--param. & $l$--param. \\
\midrule
23   & 0.067 & 504  & 57--73 & 57, 49   & 10, 6   & $4.9 \times 10^{-6}, 1.0 \times 10^{-2}$ & $2.6 \times 10^{-8}, 9.0 \times 10^{-5}$ \\
105  & 0.11 & 672  & 32--42 & 167, 143 & 12, 8   & $2.4 \times 10^{-4}, 2.6 \times 10^{-1}$ & $6.7 \times 10^{-8}, 3.5 \times 10^{-5}$ \\
510  & 0.21 & 864  & 17--24 & 430, 363 & 15, 11  & ---                            & $1.4 \times 10^{-7}, 1.2 \times 10^{-5}$ \\
1250 & 0.36 & 1152 & 10--17 & 635, 589 & 31, 16  & ---                            & $1.2 \times 10^{-6}, 7.6 \times 10^{-5}$ \\
2770 & 0.43 & 1404 & 7.2--12 & 884, 795 & 36, 21  & ---                            & $3.2 \times 10^{-6}, 5.3 \times 10^{-5}$ \\
\midrule
\multicolumn{8}{c}{Quasi-circular model}\\
\midrule
6020 & 0 & 50 & 7.2--10 & \multicolumn{2}{c}{4, 4 $(A_{22}, \phi_{22})$} & \multicolumn{2}{c}{$1.9 \times 10^{-9}, 3.1 \times 10^{-8}$} \\
\bottomrule
\end{tabular*}
\end{table*}

\begin{figure}
    \centering
    \includegraphics[width=\columnwidth, trim=0.34cm 0.35cm 0.32cm 0.33cm, clip]{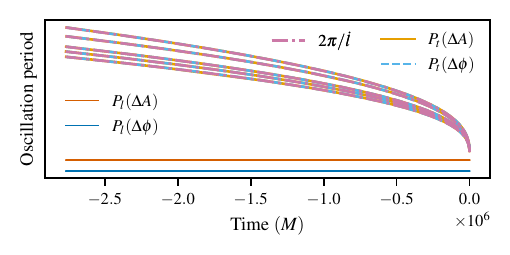}
    \caption{Variation of time/mean-anomaly period of oscillations $P_t(\Delta A)$/$P_l(\Delta A)$ and $P_t(\Delta \phi)$/$P_l(\Delta \phi)$ in $\Delta A$ and $\Delta \phi$ respectively as a function of time for a representative sample of 5 binaries. Their time period of oscillations secularly decreases as the binary inspirals (orange solid and light blue dashed curves), which is the typical chirp of a GW signal. This secular decrease is also encoded in their rate of evolution of mean anomaly angle (pink dash-dot curve; also see Eq. \eqref{eqn: mean-anomaly}), as evidenced by the mutual overlap of these three sets of curves.
    Therefore, the oscillation period of $\Delta A$ and $\Delta \phi$ in terms of the \textit{mean anomaly} for \textit{any} binary system becomes constant, i.e., $P_l(\Delta A) = P_l(\Delta \phi) = 2\pi$ (solid lines; they are vertically offset for visual clarity). In this manner, the chirping behavior of the $\Delta A$ and $\Delta \phi$ oscillations can be factored away into the non-oscillatory time evolution of mean anomaly, simplifying their harmonic content significantly.
    }
    \label{fig: chirp}
\end{figure}

\textit{Mean anomaly parameterization---}
$\Delta A$ and $\Delta \phi$ require a relatively large number of basis functions owing to their intricate harmonic content. They oscillate with binary's radial motion, and the period of these oscillations decreases secularly as the binary inspirals under gravitational radiation reaction, producing the characteristic chirping behavior (see Fig.~\ref{fig: chirp}). The chirp rate also depends on a binary's eccentricity and mass-ratio. Hence, their oscillation periods also evolve \textit{differently} across the parameter space. Accurately representing this family of oscillatory functions---whose oscillation periods vary both in time and across parameter space--- therefore requires a large number of basis functions.

To address this difficulty, we parameterize the eccentric residuals in terms of the \textit{mean anomaly} instead of time. Mean anomaly is an angular variable defined as~\citep{goldstein2002classical, Memmesheimer:2004cv},
\begin{equation}
    l(t) = \int^t n(t') \, dt'; \quad n(t) = 2 \pi/P(t), 
    \label{eqn: mean-anomaly}
\end{equation}
where $P$ is the time period of the binary's radial (periastron-to-periastron) motion, and $n$ is its \textit{orbit-averaged} radial frequency, which increases secularly as the binary inspirals. Mean anomaly thus serves as a radial orbit ``counter", advancing by $2 \pi$ with each periastron passage, while elapsing at an increasingly faster rate as the inspiral progresses, encoding the chirp.

This makes mean anomaly a particularly convenient parameterization for simplifying the structure of eccentric residuals. Expressing $\Delta A$ and $\Delta \phi$ as a function of the mean anomaly ensures that their oscillations have a \textit{constant} period of $2 \pi$ for \textit{any} ({planar}) binary configuration \textit{throughout its inspiral}. The chirp is absorbed into the time derivative of the mean anomaly (i.e., $n$; cf. Eq.~\eqref{eqn: mean-anomaly} and Fig.~\ref{fig: chirp}), rather than appearing as a varying oscillation period in the residuals. Consequently, these mean anomaly-parameterized eccentric residuals have significantly simpler harmonic content and can be represented using \textit{more than an order of magnitude fewer basis functions} than their time-parameterized counterparts (see Fig.~\ref{fig: surrogate-scaling} and Table~\ref{tab: surrogate-metrics}).

For surrogate construction, the waveform data-pieces are first represented on a common shifted mean anomaly grid, $l_s(t; \param) \coloneq l(t; \param) - l(t=0; \param)$. Then, to obtain the final waveforms as a function of time, we also make a surrogate model of the mapping between shifted mean anomaly and time (see Appendix A for details). Mean anomaly is a monotonically increasing, \textit{non-oscillatory} function of time (see Fig. \ref{fig: chirp}, which shows its time-derivative) because it evolves via the \textit{orbit-averaged} radial orbital frequency (cf. Eq. \eqref{eqn: mean-anomaly}), and a highly compressed surrogate can be easily built for it.

Angular element representations have proven useful in extreme mass-ratio inspiral (EMRI) modeling as well, from avoiding numerical issues near orbital turning points~\citep{Cutler:1994pb, Glampedakis:2002ya} to waveforms that are harmonically-decomposed in orbit-averaged frequencies and angles~\citep{Pound:2021qin, Drummond:2022xej, Drummond:2022efc, VanDeMeent:2018cgn, Lynch:2023gpu, Drummond:2023wqc, Chua:2020stf, Katz:2021yft, Speri:2023jte, Chapman-Bird:2025xtd}. However, our framework is the first to leverage the mean anomaly domain to build efficient surrogate models of eccentric waveforms.


\textit{Simplification of parameter space fits---} 
To complete the surrogate model, one constructs fits for each data piece's values at a set of time/shifted mean anomaly nodes obtained via the Empirical Interpolation (EI) method~\citep{Field:2013cfa}.
We find that for \textit{both} time and mean anomaly-parameterized surrogates, the variations in $\Delta A$ and $\Delta \phi$ are significantly simplified when expressed against the \textit{instantaneous} values of eccentricity and mean anomaly $(e_{EI}, l_{EI})$ at the corresponding EI time/shifted mean anomaly nodes, instead of their values $(\eref, \lref)$ at a fixed reference time $t_{\rm{ref}}$ (see Fig.~\ref{fig: parameter-space-variation}).
This allows accurate parametric fits to be built from a relatively sparse training waveform space. We detail the surrogate construction steps in Appendix A. For the same reason, we parameterize the surrogate via the symmetric mass ratio ($\eta \coloneq m_1 m_2/(m_1+m_2)^2$) instead of the mass-ratio $(q)$ for all data-pieces.

\textit{Results---} We construct surrogates of increasing waveform durations using both the conventional time-parameterized, as well as our novel mean anomaly-based approaches. Their details are summarized in Table~\ref{tab: surrogate-metrics} for comparison.
The mean anomaly-parameterized data-pieces require \textit{an order of magnitude fewer basis functions}, with a milder scaling of basis size with surrogate length (cf. Fig.~\ref{fig: surrogate-scaling}). 
To assess surrogate accuracy, we compute mismatches against $10,000$ waveforms from the base \inspiralesigma{} model randomly sampled across the surrogate parameter space. The mismatches are computed over the full surrogate duration using the zero detuning high power noise spectral density for LIGO detectors~\citep{LIGO-T0900288-v3, LIGO-T070247-v1} for a $10 M_\odot$ binary. With highest mismatches of $\mathcal{O}(10^{-5})$, the mean anomaly-parameterized surrogates are substantially more accurate than the time-parameterized surrogates, which exhibit a tail of high mismatches reaching $\mathcal{O}(10^{-1})$ (cf. Table~\ref{tab: surrogate-metrics}).
These results demonstrate the superior scalability and accuracy of the mean anomaly-parameterized approach for constructing long-duration eccentric surrogate models.

\begin{figure*}
    \centering
    \includegraphics[width=\textwidth, trim=0.24cm 0.3cm 0.25cm 0.25cm, clip]{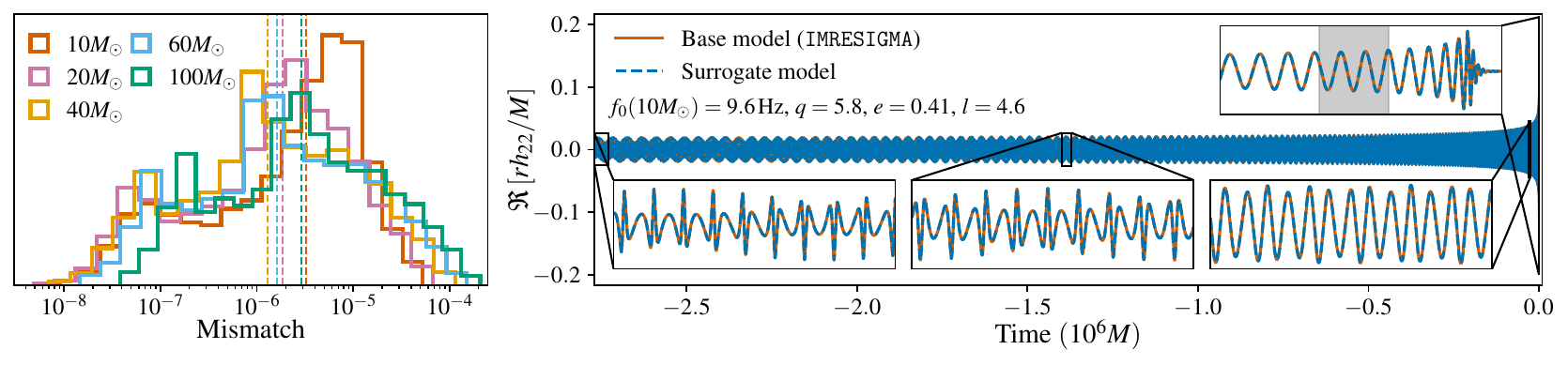}
    \caption{Accuracy of the $2.77 \times 10^6M$ long mean anomaly-parameterized inspiral surrogate \inspiralesigmasur{} (cf. Table~\ref{tab: surrogate-metrics}). 
    \textit{Left:} Mismatches against \inspiralesigma{} for $10,000$ binaries sampled across the surrogate parameter space at different total masses; dashed lines show the median values. 
    \textit{Right:} Illustration of \inspiralesigmasur{} used as a drop-in replacement of \inspiralesigma{} within the \esigma{}~\citep{Paul:2024ujx, esigmapy} framework to produce complete IMR waveforms. The (orange) \imresigma{} waveform is constructed by smoothly attaching the long inspiral piece (\inspiralesigma{}) to a short plunge-merger-ringdown (PMR) piece over a short time-window (gray band in the top right inset).  The blue curve shows an IMR waveform constructed similarly, utilizing \inspiralesigmasur{} instead of raw \inspiralesigma{}. The parameters at the beginning of the waveform correspond to the case yielding the worst inspiral surrogate mismatch ($5.3 \times 10^{-5}$) against \inspiralesigma{} at $10M_\odot$. 
    As highlighted via the insets, \inspiralesigmasur{} remains faithful to the base \inspiralesigma{} model throughout.}
    \label{fig: mismatch-IMR-surrogate}
\end{figure*}

Utilizing the scalability of our approach, we construct \inspiralesigmasur{}: a $2.77 \times 10^6M$ ($850-1250$ orbits) long non-spinning eccentric inspiral surrogate with a maximum starting eccentricity of $0.43$. Depending on the binary parameters, this time duration corresponds to a $10M_\odot$ binary starting with a GW frequency of $7.2-12$Hz. The left panel of Figure~\ref{fig: mismatch-IMR-surrogate} shows its mismatches against \inspiralesigma{} for a population of eccentric binaries. It achieves median mismatches of $1.3-3.2 \times 10^{-6}$, with the worst mismatch of $2.1 \times 10^{-4}$ for a $100M_\odot$ system. Thus, \inspiralesigmasur{} is faithful to its base model \inspiralesigma{} and can serve as its drop-in replacement within the \esigma{} framework~\citep{Paul:2024ujx, esigmapy} for producing complete IMR eccentric waveforms (see the right panel of Fig.~\ref{fig: mismatch-IMR-surrogate} for an illustration).

\begin{figure}
    \centering
    \includegraphics[width=\columnwidth, trim=0.35cm 0.3cm 0.35cm 0.3cm, clip]{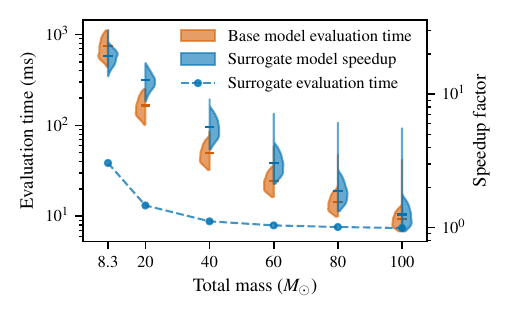}
    \caption{Waveform evaluation time of the base model \inspiralesigma{} (orange distribution), and the corresponding speedup (blue distribution) and the median evaluation time (blue dots) of its $2.77 \times 10^6M$ (850-1250 orbits) long mean anomaly-parameterized surrogate \inspiralesigmasur{} for different binary masses. The waveforms are generated from $15$Hz at a sampling rate of $4096$Hz at 1000 points randomly drawn across the surrogate parameter space for each binary mass. The lowest total mass shown corresponds to the smallest value for which the surrogate can be evaluated across its entire parameter space from $15$Hz. Markers also indicate the respective median evaluation times/speedup. The study was performed on an AMD EPYC 7352 processor operating at 2.3 GHz.
    }
    \label{fig: speedup}
\end{figure}

Figure~\ref{fig: speedup} compares the waveform generation time of \inspiralesigmasur{} with \inspiralesigma{}. With a starting GW frequency of $15$Hz and a sampling rate of $4096$Hz, \inspiralesigmasur{} is about $20$ times faster to evaluate at its lowest allowed total mass of $8.3M_\odot$, with an evaluation time of about $40$ms, thus making it adequate to analyze such light binary systems using routine LVK Bayesian parameter estimation pipelines.  
%

\textit{Conclusions---}
In this \textit{Letter}, we presented a novel, uniquely scalable technique for constructing surrogates for very long-duration eccentric BBH waveforms. The technique simplifies the harmonic content of eccentricity-induced oscillations in the waveform data-pieces by parameterizing them in terms of the \textit{mean anomaly} angle instead of time. This enables construction of an order of magnitude more compressed surrogate models than all contemporary time-parameterized methods~\citep{Field:2013cfa, Islam:2021mha}, while being orders of magnitude more faithful to the base waveform model. We also significantly simplify the parameter-space fitting of these oscillatory waveform data-pieces by parameterizing them against the \textit{instantaneous} orbital eccentricity and mean anomaly values. 

We demonstrate the scalability of this scheme by constructing \inspiralesigmasur{}: a $2.77\times 10^6M$ ($850-1250$ orbits) long non-spinning, eccentric surrogate of the waveform model \inspiralesigma{}~\citep{Paul:2024ujx} with starting eccentricities up to $0.43$. This can be used to analyze BBHs of masses as low as $8.3M_\odot$ from $15$Hz. The surrogate is efficient, taking about $40$ms to evaluate at its full duration, speeding up the waveform generation by an order of magnitude. It can be accessed via the Python package \esigmapy{}~\citep{esigmapy, esigmasur-zenodo}. The scalability of this technique will prove useful for constructing eccentric surrogates not only for current GW detectors but also for the \textit{future third-generation detectors that will require orders of magnitude longer waveform templates} to observe nearly all the BBH mergers within the era of star formation~\citep{Reitze:2019iox, ET:2019dnz}.

While we have developed surrogates for \inspiralesigma{}, the techniques presented here are general.
In particular, using definitions of eccentricity and mean anomaly based solely on waveform morphology~\citep{Mora:2002gf, Shaikh:2023ypz, Boschini:2024scu, Islam:2025oiv, Shaikh:2025tae} should make this method agnostic to the internal details/conventions of any particular waveform model. 
Similarly, while our focus here has been to build scalable long-duration eccentric surrogates, Ref. \cite{Nee:2025nmh} adapts this technique to construct accurate short-duration \textit{eccentric numerical relativity surrogate models}. 
The method should also be extendable to higher eccentricities, likely requiring modestly more basis functions to capture the sharper eccentric features. This is particularly relevant given the impact of eccentricity on sky localization and rapid electromagnetic follow-up~\citep{Pan:2019anf, Yang:2022fgp, Pal:2025iyo, Sinha:2025vmc}.

Work is ongoing to extend the framework to aligned-spin BBH systems and to include higher-order GW modes beyond the dominant (2,2)-mode. Unlike the ODE-integrated waveform models that are bound to generate waveforms serially, the operations during surrogate evaluation are highly amenable to parallelization and hardware acceleration (e.g., \cite{Khan:2020fso, Thomas:2022rmc, Thomas:2025rje, Edwards:2023sak}) and we are also working in this direction to further accelerate waveform generation.

\begin{acknowledgments}
We would like to thank Md. Arif Shaikh and Scott Hughes for useful comments on the work. We also acknowledge journal referees for their constructive feedback. We thank the members of the Astrophysics \& Relativity group at the International Centre for Theoretical Sciences (ICTS) for valuable discussions. A.M. and P.K. acknowledge support of the Department of Atomic Energy, Government of India, under project no. RTI4001 at ICTS. A.M. is an Infosys Fellow supported by the Infosys Foundation and acknowledges the financial support from the Infosys-TIFR Leading Edge travel grant. P.K. acknowledges the support by the Ashok and Gita Vaish Early Career Faculty Fellowship at ICTS.  
CKM acknowledges the support of SERB’s Core Research Grant No. CRG/2022/007959.
V.V.~acknowledges support from NSF Grant No. PHY-2309301.
S.F. acknowledges support from NSF Grants No. AST-2407454 and PHY-2110496.
Parts of this work were carried out during visits to the Institute for Computational and Experimental Research in Mathematics (ICERM) at Brown University; A.M. and P.K. gratefully acknowledge their financial support, with A.M. also acknowledging the support from the International Travel Support (ITS) scheme under the Anusandhan National Research Foundation (ANRF). A.M. also acknowledges the hospitality of the Max Planck Institute for Gravitational Physics (Albert Einstein Institute), Potsdam, where a part of this work was completed.
This work was partly supported by UMass Dartmouth's Marine and Undersea Technology (MUST) research program funded by the Office of Naval Research (ONR) under grant no. N00014-23-1-2141.
All the computations for this work were carried out on the Sonic computing cluster at ICTS. The LIGO document number of this work is LIGO-P2500585. 
\end{acknowledgments}

\section*{Data Availability}
The surrogate data files for \inspiralesigmasur{} can be found at~\cite{esigmasur-zenodo}, and they can be used to evaluate it through the Python package \esigmapy{}~\citep{esigmapy}.  

\bibliography{References.bib}

\appendix
\onecolumngrid
\section*{End Matter}
\twocolumngrid

\textit{Appendix A: Construction of mean-anomaly domain surrogate for BBH inspirals---}
Collectively denoting the eccentric residuals $\{\Delta A$, $\Delta \phi$, $\phi_{\rm{res}}\}$ by $\bm{f}$, we list the steps of constructing the mean anomaly-parameterized surrogate below. 
\begin{enumerate}
    \item Assuming a training set of $N_{\rm{train}}$ inspiral-only waveforms with parameters $\{\param_i\}_{i=1}^{N_{\rm{train}}}$, we extract the \textit{unwrapped} mean anomaly evolution as a function of time $l(t; \param_i)$ from the waveform model's orbital dynamics solver~\footnote{It should also be possible to work with a definition of mean anomaly that only depends on the waveform morphology~\citep{Shaikh:2023ypz, Shaikh:2025tae}, making this method agnostic to the internal details/conventions of any particular eccentric waveform model; we leave this for future work.}.
        
    We also make sure that the \textit{unwrapped} $l(t; \param_i)$ of each waveform has its \textit{wrapped} value of $l_{{\rm{ref}}_i}$ at $t_{\rm{ref}}$. This avoids discontinuous $2 \pi$-jumps in the variation of $l(t; \param)$ with $\param = \lbrace \eref, \lref, \bm{\theta'} \rbrace$, despite its inherent $2 \pi$-periodicity/degeneracy.
    
    \item We find the mean-anomaly extent of each waveform $\Delta l(\param_i) = l(t=0; \param_i) - l(t=t_0; \param_i)$, where $t_0$ is the starting time of the particular waveform. We then calculate the maximum possible common mean anomaly extent $L$ of the training space waveforms by calculating the minimum of their extents, i.e. $L = \mathrm{min}_i\, \Delta l(\param_i)$. We also denote by $T_L(\param_i)$ the time durations of the waveforms having a mean anomaly extent of $L$. 
    \item Our waveforms end at a constant orbit-averaged orbital frequency at $t=0$, so they generally end at different mean anomaly values. For surrogate construction, however, it is necessary to represent the data-pieces on a common grid. We therefore model the data-pieces against the \textit{shifted mean anomaly},   
    \begin{equation}
        l_s(t; \param) = l(t; \param) - l(t=0; \param).
        \label{eqn: shifted mean anomaly}
    \end{equation}
     This ensures that all the waveforms terminate at $l_s=0$. The shifted mean anomaly, being related to the bare mean anomaly by a constant offset, still has the same time derivative and hence continues to factor away the chirp from the data-pieces (cf. Fig. \ref{fig: chirp}). We express all the eccentric residuals ($\bm{f}$) against it, i.e., as $\bm{f}(l_s; \param_i)$ in $l_s \in [-L, 0]$, and make their surrogates $\bm{f_s}(l_s; \param)$. We set $\phi_{22}(l_s=-L; \param)=\phi_{22}(l_s=-L; \eref=\lref=0, \bm{\theta'})=0$, thus ensuring that $\Delta \phi_1(l_s=-L; \param)=0$ (cf. Eq. \ref{eqn: residual-phase}).
    \item We also construct a surrogate $l_s^{\rm{sur}}(t; \param)$ of the shifted mean anomaly as a function of time for $t \in [-T_{\rm{min}}^L, 0]$, where $T_{\rm{min}}^L = \mathrm{min}_i \, T_L(\param_i)$ is the maximum possible surrogate length in time for a training space with mean-anomaly extent $L$. Since $l_s$ is a monotonically increasing, non-oscillatory function of time (see Fig. \ref{fig: chirp}), its surrogate requires only a few $(<10)$ basis functions.
    \item Using the shifted mean anomaly surrogate, we can get the surrogates of the eccentric data-pieces as a function of time: $\bm{f_s}(t; \param) = \bm{f_s}(l_s=l_s^{\rm{sur}}(t; \param); \param)$, where $t \in [-T_{\rm{min}}^L, 0]$.
    \item For \textit{both} time-parameterized and mean anomaly-parameterized surrogates, variations in the eccentric residuals $\Delta A$ and $\Delta \phi$ across the parameter space at a particular EI node are significantly simplified when parameterized by the instantaneous values of eccentricity ($e_{\rm{EI}}^k$) and mean anomaly ($l_{\rm{EI}}^k$) at the EI nodes (labeled by the index $k$), instead of their values $(e_{\rm{ref}}, l_{\rm{ref}})$ at a fixed reference time $t_{\rm{ref}}$. This is illustrated in Fig. \ref{fig: parameter-space-variation} for mean anomaly-parameterized $\Delta A$ at a shifted mean anomaly EI node. However, in the mean anomaly-parameterized case, we find that the parameter space variations in $\phi_{\rm{res}}$ at the shifted mean anomaly EI nodes show oscillations. We found that fixing the eccentric residual phase and the monotonic trend initially to zero, i.e., $\Delta \phi(l_s=-L)=\phi_{\rm{res}}(l_s=-L)=0$ simplifies their structure. This choice still respects the zero starting $(2,2)$-mode phase condition $\Delta \phi_1(l_s=-L; \param)=0$. Hence, we use these initially zeroed-out $\Delta \phi$ and $\phi_{\rm{res}}$ for the mean anomaly-parameterized surrogate.

    We use Gaussian Process Regression (GPR) for building the parametric fits \cite{Varma:2018aht}. However, due to the high evaluation cost of GPR fits, we replaced them with faster tensor product cubic spline interpolants~\cite{TPI} constructed over the GPR fit predictions.

    \begin{figure}
        \centering
        \includegraphics[width=\columnwidth, trim=0.35cm 0.125cm 0.2cm 0.8cm, clip ]{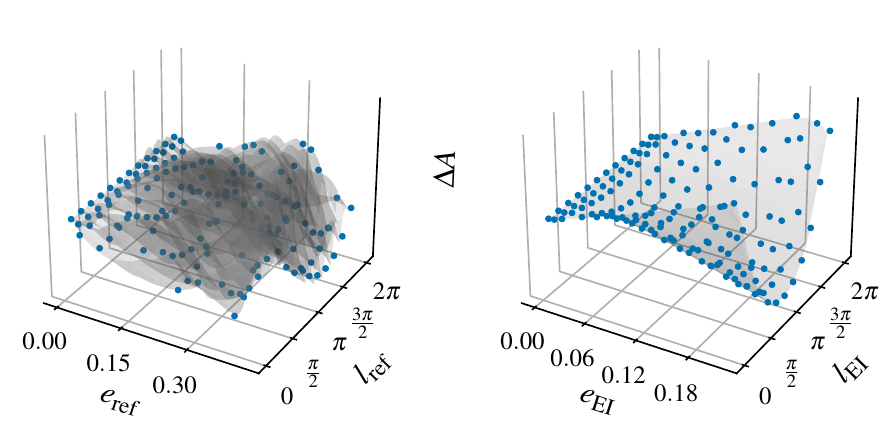}
        \caption{Variation of the eccentric residual amplitude $\Delta A$ at a single empirical interpolation (EI) node across the parameter space for a fixed mass ratio $(q = 3.2)$ for the $2.77 \times 10^6M$ long surrogate. The parameterization in terms of instantaneous values of eccentricity ($e_{\rm{EI}}$) and mean anomaly ($l_{\rm{EI}}$) at the EI nodes (right) yields smoother variation than a parameterization in terms of their values $e_{\rm{ref}}$ and $l_{\rm{ref}}$ at a fixed reference time $t_{\rm{ref}}$ (left). The simplified structure allows for the construction of accurate parametric fits with relatively fewer training space points. $\Delta A$ at a shifted mean anomaly EI node is shown here, and a similar simplification is obtained at time EI nodes as well. 
        }
        \label{fig: parameter-space-variation}
    \end{figure}

    \item Lastly, we build surrogates of the eccentricity $e(l_s; \param)$ and the (unwrapped) mean anomaly $l(l_s; \param)$ evolution as a function of $l_s$, to evaluate the values of eccentricity ($e_{\rm{EI}}^k$) and (wrapped) mean anomaly ($l_{\rm{EI}}^k$) at the shifted mean anomaly EI nodes $l_s=L_s^k$ at any $\param$ \footnote{Similarly, we build surrogates of the eccentricity $e(t; \param)$ and the (unwrapped) mean anomaly $l(t; \param)$ evolution as a function of $t$ for the time-parameterized surrogates constructed in this work.}. However, since $l_s(t; \param) = l(t; \param) - l(t=0; \param)$, we simply build a parameter space fit for $l(t=0; \param)$ to get the mean anomaly at EI nodes as $l(l_s=L_s^k; \param) = L_s^k + l(t=0; \param)$.
\end{enumerate}
\end{document}